\documentclass[a4paper,11pt]{article}
\usepackage{pos}
\usepackage{tabularx}
\usepackage{lineno}
\title{A time-independent search for neutrinos from galaxy clusters with IceCube}
 \ShortTitle{IceCube Galaxy Clusters}

\author{The IceCube Collaboration \\{\normalsize \normalfont(a complete list of authors can be found at the end of the proceedings)}}
\emailAdd{mnisa@icecube.wisc.edu}
\emailAdd{aludwig@icecube.wisc.edu}

\abstract{Clusters of galaxies — with their turbulent magnetic fields and abundant matter content — are a promising class of potential neutrino sources. Cosmic rays accelerated within the large-scale shocks, Active Galactic Nuclei (AGN), or both can be confined in galaxy clusters over cosmological timescales and produce a steady flux of neutrinos in secondary interactions. The IceCube Neutrino Observatory has detected a diffuse flux of high-energy astrophysical neutrinos. After ten years of operations, however, the origin of this flux remains largely unconstrained. In this work, we perform a stacked search for neutrinos, using a population of over one thousand galaxy clusters detected by the Planck Satellite via the Sunyaev-Zeldovich (SZ) effect up to a redshift $z = 1$. We present the first results on the contribution of galaxy clusters to the diffuse neutrino flux and discuss the implications for various models of cosmic-ray acceleration in large-scale structures.

\vspace{4mm}
{\bfseries Corresponding authors:}
Mehr Un Nisa$^{1*}$, Andrew Ludwig$^{1}$\\
{$^{1}$ \itshape Michigan State University}\\
$^*$ Presenter
\FullConference{37$^{\rm{th}}$ International Cosmic Ray Conference (ICRC 2021)\\
		July 12th -- 23rd, 2021\\
		Online -- Berlin, Germany}

}

\begin{document}
\maketitle

\section{Introduction}\label{sec:intro}

Despite recent advances in multimessenger astronomy and first indications of potential sources of astrophysical neutrinos ~\cite{txs,txs2,Aartsen:2019fau}, the origin of the majority of the high-energy neutrino flux observed by IceCube remains shrouded in mystery. 
Galaxy clusters are a potential class of sources that have mostly remained uninvestigated by IceCube thus far. Previous IceCube studies have constrained neutrino emission from only five nearby galaxy clusters
~\cite{IC40_cluster,IC40_cluster2}.
As with other source classes, galaxy clusters are expected to produce neutrinos as secondary particles, through the acceleration and interaction of cosmic rays (CRs). 

There are two major proposed mechanisms for CR acceleration in clusters.
The first mechanism involves gas accretion during large-scale structure formation, which can produce megaparsec (Mpc) scale shocks~\cite{miniati2000,ryu2003} that can accelerate cosmic rays to high energies.
The second mode of CR acceleration involves sources embedded within clusters, such as active galactic nuclei (AGN)~\cite{stecker1991,winter2013}.
In this work we treat galaxy clusters as steady sources of neutrinos.

The accelerated CRs are then confined by the magnetic fields of the clusters, which can trap CRs for times exceeding the age of the universe~\cite{berezinsky1997}, giving them ample time to participate in proton-proton interactions with the intra-cluster medium (ICM).
These interactions produce pions, which decay, producing neutrinos and $\gamma$ rays. The class of models that predict neutrino fluxes from the aforementioned mechanisms are also referred to as ``CR reservoir'' models, because the ICM acts as a reservoir of CRs. This work tests a few such models that predict a neutrino flux potentially detectable by IceCube~\cite{fangolinto2016,fangmurase2018}. 

\section{Galaxy Cluster Catalog}
\label{sec:catalog}

We use the catalog of galaxy clusters from the Planck 2015 survey~\cite{planck_2015}.
Galaxy clusters were observed by Planck through the Sunyaev–Zeldovich (SZ) effect over the course of a 29 month mission.
We include all clusters with mass and redshift information, for a total of 1094 sources.
Because we are using a catalog that includes a wide variety of masses and redshifts, and we expect neutrino emission to depend on these properties, it is useful to weight our sources in various ways, detailed in Section~\ref{sec:catalog:w}.

\subsection{Catalog Completeness}

In order to convert the sensitivity estimates for this catalog into an estimate for the entire population of galaxy clusters within the redshift range accessible via the SZ effect, we need an estimate of the completeness of the catalog. The percentage of completeness can be estimated by comparing a theoretical distribution of galaxy clusters above a given mass threshold with the completeness function for the Planck survey. 
Planck's completeness is a measure of how reliably Planck expects to detect galaxy clusters, as a function of mass and redshift of the cluster.
We extract Planck's completeness function directly from Ref.~\cite{planck_2015}.
With this completeness function in hand, we also need an expected distribution of galaxy clusters. For this we sample clusters from the Tinker 2010 halo mass function~\cite{tinker_2010}.
The total catalog completeness is estimated by convolving together the halo mass function and Planck's completeness function. 
We only consider galaxy clusters with masses between $10^{14}$ \(\textup{M}_\odot\) and $10^{15}$ \(\textup{M}_\odot\) and at a redshift between 0.1 and 2.
The estimate of catalog completeness depends on how the galaxy clusters are weighted.
In the case of all of these clusters weighted equally, we find a completeness of $\sim 17 \%$.

\subsection{Weighting}
\label{sec:catalog:w}

\begin{figure}[ht!] 
    \centering
    \includegraphics[width=1.\linewidth]{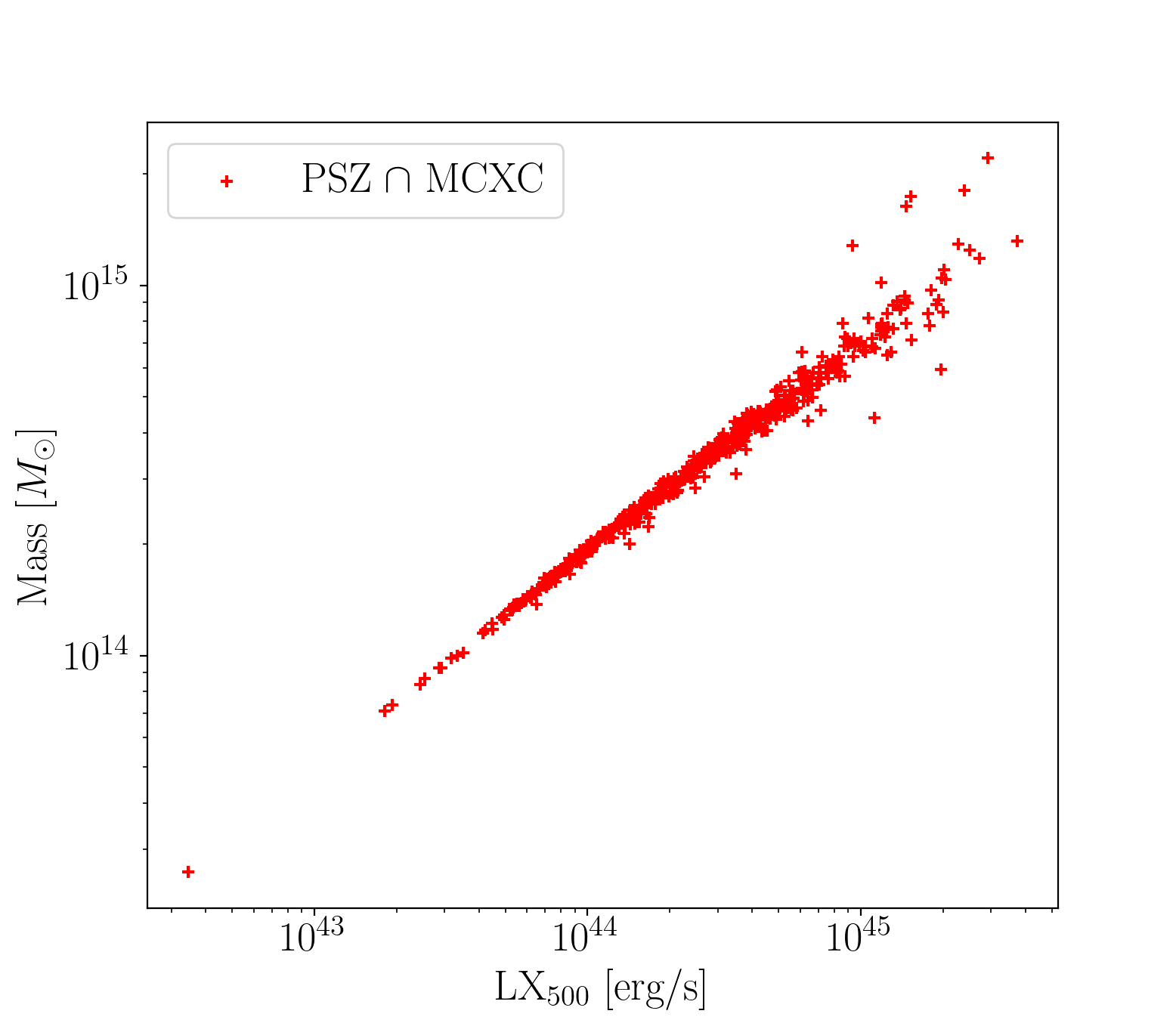}
    \caption{\label{fig:fig1} The standardised 0.1 -- 2.4 keV band X-ray luminosity, $Lx_{500}$,  and the mass of clusters, for the 518 clusters in the Planck catalog that have a counterpart in the MCXC catalog \cite{mcxc} of X-ray detected clusters. The X-ray luminosity is highly correlated with the mass of clusters. We use the cluster mass as a proxy as the X-ray luminosity is not available for all clusters.}
\end{figure}

We expect that the contributions of galaxy clusters to the diffuse neutrino flux depend on their mass and redshift, so we also derive an ``effective completeness'' by weighting according to these properties.
The three weighting schemes we consider in this work are equal weights, inverse distance-squared weighting, and mass times inverse distance-squared weights. The weight of each cluster $k$ is denoted by $w_{k}$.  
Equal weights translate as $w_{k} = 1$ for all $k$, effectively assigning equal probability for neutrino emission to each candidate source, regardless of any of its properties.
Inverse distance-squared weighting assigns weights as $w_{k} = \frac{1}{d_{k}^{2}}$, where $d_{k}$ is the distance to a given cluster.
This type of weighting assumes that, regardless of a source's other properties, the neutrino flux will fall off as a function of the distance.
We refer to this as distance weighting throughout this work.
The mass times inverse distance-squared scenario weights sources by $w_{k} = \frac{M_{k}}{d_{k}^{2}}$, where $M_{k}$ is the source mass. 
The clusters in the Planck survey that have an X-ray counterpart show a high degree of correlation between the mass of the cluster and its X-ray luminosity -- which in turn measures the ICM density and temperature (Figure \ref{fig:fig1}). 
Therefore, this weighting scheme implicitly assumes that the neutrino luminosity is proportional to the galaxy cluster mass, making this effectively a proxy for the neutrino flux, and as such we refer to this as flux-proxy weighting throughout this work. 
These three weighting cases have estimated effective completenesses of $\sim17\%$, $\sim 53 \%$, and $\sim 66 \%$, respectively.

%%%%%%%%%%%%%%%%%%%%%%FIGURE%%%%%%%%%%%%%%%%%%%%%%%%%%%%

%%%%%%%%%%%%%%%%%%%%%%FIGURE%%%%%%%%%%%%%%%%%%%%%%%%%%%%

\section{Detector and Dataset}
\label{sec:detector}

The IceCube Neutrino Observatory is a cubic kilometer of South Pole ice, instrumented with 5160 digital optical modules (DOMs) placed 1450 - 2450~m below Antarctica's surface.
IceCube consists of 86 vertical strings, with 60 DOMs on each.
DOMs consist of a photomultiplier tube (PMT) and on-board readout electronics designed to detect Cherenkov light produced by secondary particles produced in neutrino interactions~\cite{icecube_detector,icecube_daq}.
Photons from these secondary interactions are recorded by DOMs, and the photon counts and timing information is used to reconstruct both energy deposition and particle direction.
IceCube can detect all flavors of neutrinos, but achieves very different angular resolution for each flavor.
Muon neutrinos are characterized by track-like signatures in the detector, with an angular resolution typically below 1$^\circ$ above 1 TeV ~\cite{icecube_7year}.

This analysis uses a 9.5 year data sample consisting primarily of muon neutrino tracks due to the better angular resolution.
The dataset covers the full sky, recorded between April 2008 and November 2017~\cite{icecube_7year,realtime}.  
Most of the events in the dataset are background atmospheric neutrinos and muons created in cosmic ray showers.
There are six distinct periods of detector development represented in this dataset, as in 2008, there were only 40 IceCube strings, and the full 86-string configuration was completed in 2012.
In addition to the changing string number, event selections and data-taking conditions have changed over the years and are also accounted for.

\section{Analysis Methods}
\label{sec:analysis}

With the dataset of tracks and the catalog of galaxy clusters in hand, we perform a stacking search with an unbinned maximum-likelihood method.
Sources are stacked with the standard unbinned point-source likelihood detailed in~\cite{unbinned_llh}, with a stacking technique similar to~\cite{stacking}.
The function for this unbinned likelihood method for $M$ stacked clusters is defined as:
\[ 
\mathcal{L} \left( n_{s}, \gamma \right) = \prod_{i}^{N} \left( \frac{n_{s}}{N} \sum_{k}^{M} w_{k} S_{i}^{k} + \left( 1 - \frac{n_{s}}{N} \right) B_{i} \right),
\]
where $n_{s}$ is the number of signal events, $\gamma$ is the spectral index of an assumed unbroken power-law spectrum for neutrino flux that is common to all sources, $N$ is the total number of data events, $S_{i}^{k}$ is the signal probability distribution function (PDF) for the $i$th data event for the $k$th source, $B_{i}$ is the background PDF for the $i$th data event, and $w_{k}$ is the source weight. 

Both the signal and background PDFs are functions of spatial coordinates and energy.
The background PDF is a function of both the event energy proxy, $E_{i}$, and the reconstructed declination, $\delta_{i}$.
The signal PDF has a spatial term and an energy term, with an assumed Gaussian term for the spatial part, and a simple power-law for the energy part.
A signal PDF's spatial term depends on $\delta_{i}$, the reconstructed right ascension, $\alpha_{i}$, and the angular uncertainty of the event, $\sigma_{i}$, while the energy term depends on $E_{i}$ and an overall spectral index for all sources, $\gamma$, which is assumed to have a value between 1 and 4. 
%The signal is simulated with an assumed Gaussian term for the spatial part, and a simple power-law for the energy part.

Our test statistic (TS) is twice the ratio of the log of the ratio of the best-fit likelihood to the null (background-only) hypothesis:
\[
\text{TS} = 2\log\frac{\mathcal{L}\left( \hat{n}_{s}, \hat{\gamma}\right)}{\mathcal{L} \left( n_{s} = 0 \right)},
\]
where $\hat{n}_{s}$ and $\hat{\gamma}$ are the best fit values.
We perform a blind analysis by first randomizing the neutrino data in right ascension and running the maximum-likelihood method many times to create a TS distribution that is approximately $\chi^{2}$ distributed~\cite{wilks}.
%Using this TS distribution, we can approximate our sensitivity and discovery potential.
The TS value for our real (unrandomized) data is compared to our background TS distribution after unblinding, which gives us a $p$ value for how compatible our data is with the null hypothesis.

\section{Results}
\label{sec:results}
%%%%%%%%%%%%%%%%%%%%%%FIGURE%%%%%%%%%%%%%%%%%%%%%%%%%%%%

\begin{figure} 
    \centering
    \includegraphics[width=1.\linewidth]{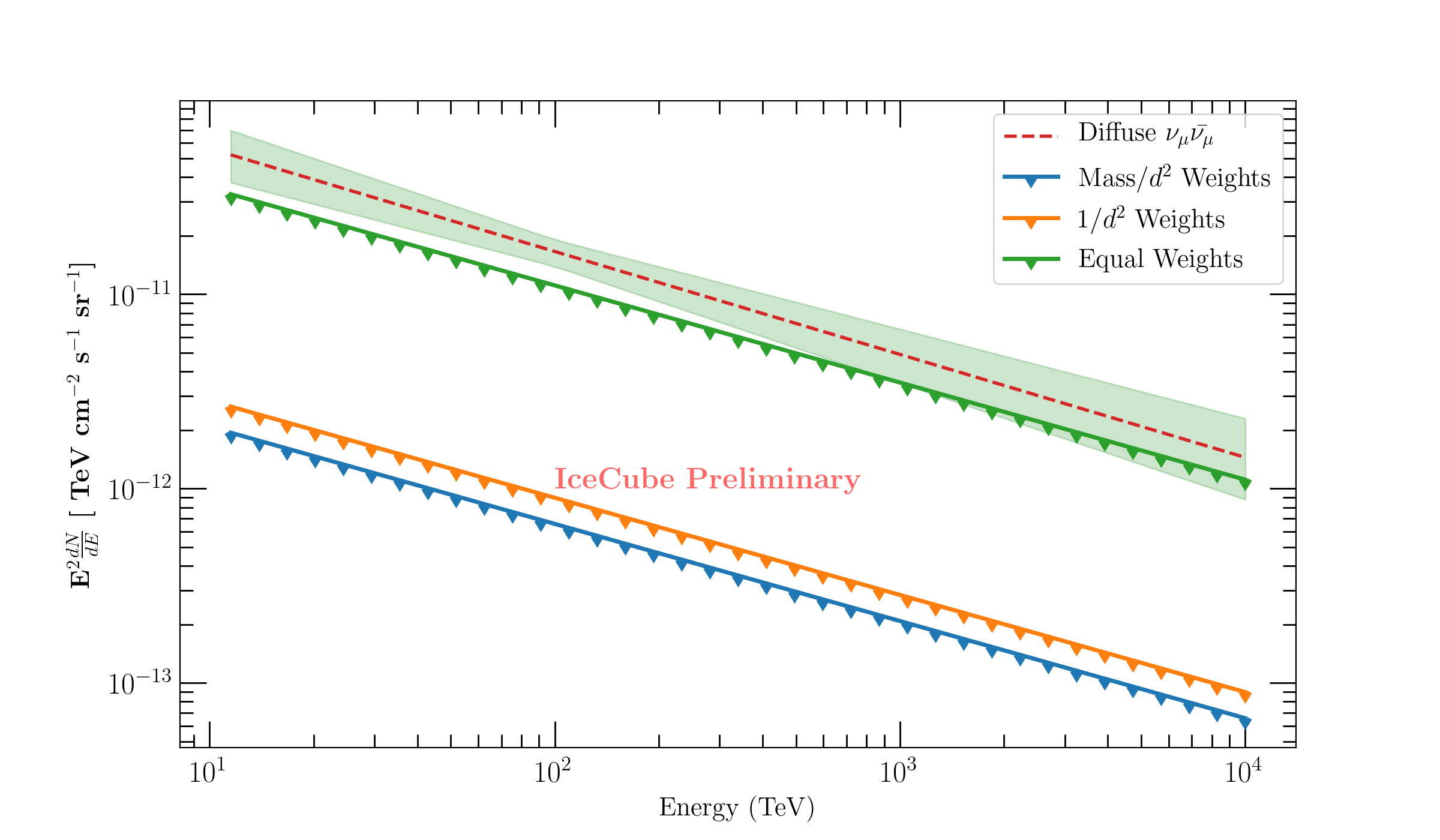}
    \caption{\label{fig:fig2} IceCube 90\% CL limits on the stacked flux from the Planck SZ Catalog of Galaxy Clusters assuming a spectral index of 2.5. The limits have been scaled for the completeness of the catalog for each weighting scheme. The fraction of the diffuse flux~\cite{diffuse} constrained at 100 TeV assuming an $E^{-2.5}$ power-law is 66.9\%, 5.3\% and 3.9\% for the unweighted, distance weighted and flux-proxy scenarios respectively.}
\end{figure}
%%%%%%%%%%%%%%%%%%%%%%FIGURE%%%%%%%%%%%%%%%%%%%%%%%%%%%%
We performed the unbinned likelihood stacking analysis as described in Section~\ref{sec:analysis} for the three different weighting schemes described in Section~\ref{sec:catalog:w}. 
None of the scenarios we investigated show significant excess of neutrino events over background.
Because all scenarios are compatible with the null hypothesis, we report 90\% CL upper limits for the total flux of neutrinos from our selection of 1094 galaxy clusters for each weighting scheme and assumed spectral indices of 2.0, 2.5, and 3.0 in Table~\ref{tab:sens}. 
Figure ~\ref{fig:fig2} shows the upper limit on the neutrino emission from clusters assuming a simple power-law flux with a spectral index of 2.5. 

The upper limits can be compared with the diffuse $\nu_{\mu} + \bar{\nu}_{\mu}$ flux measured by IceCube to obtain an estimate of the maximal contribution of the population of galaxy clusters under investigation at a given energy. 
After accounting for the completeness of the survey for each weighting scheme, we find that at 100  TeV for a spectral index of 2.5, galaxy clusters in our considered mass and redshift range can produce no more than 67\% of the diffuse flux under the equal weights scenario, which we consider the most conservative and model independent of the three scenarios.

\begin{table*}
\centering
\begin{tabularx}{\textwidth}{| X | X | X | X | X | X |}
\hline
 Weighting Scheme & TS & $n_{s}$ & Flux Limit $\gamma = 2.0$ & Flux Limit $\gamma = 2.5$ & Flux Limit $\gamma = 3.0$ \\
 \hline
 Equal & 0 & 0 & $9.38 \times 10^{-12}$ & $1.96 \times 10^{-10}$ & $8.30 \times 10^{-10}$  \\
 Distance & 0 & 0 & $3.64 \times 10^{-12}$  & $4.99 \times 10^{-11}$ & $1.34 \times 10^{-10}$ \\
 Flux-Proxy & 0 & 0 & $4.25 \times 10^{-12}$ & $4.56 \times 10^{-11}$ & $1.53 \times 10^{-10}$ \\
 \hline
 \end{tabularx}
 \caption{The results of the stacking analysis with the TS and the best-fit number of signal events $n_s$ for all three weighting schemes. The 90\% CL limits on the flux normalization at 1 TeV (before scaling for the completeness of the catalog) are shown for three different power-law assumptions. All fluxes are in units of TeV cm$^{-2}$ s$^{-1}$. }
\label{tab:sens}
\end{table*}

In our most realistic scenario, which is the flux-proxy weighting, for an assumed $\gamma = 2.5$ unbroken power-law spectrum, we find that galaxy clusters can contribute no more than $4\%$ of the diffuse flux at 100~TeV.
Table~\ref{tab:contrib} contains the 90\% CL upper limit contribution to the diffuse flux for all weighting schemes and tested power-law indices at 100~TeV.

%%%%%%%%%%%%%%%%%%%%%%FIGURE%%%%%%%%%%%%%%%%%%%%%%%%%%%%

\begin{figure}[ht!] 
    \centering
    \includegraphics[width=1.\linewidth]{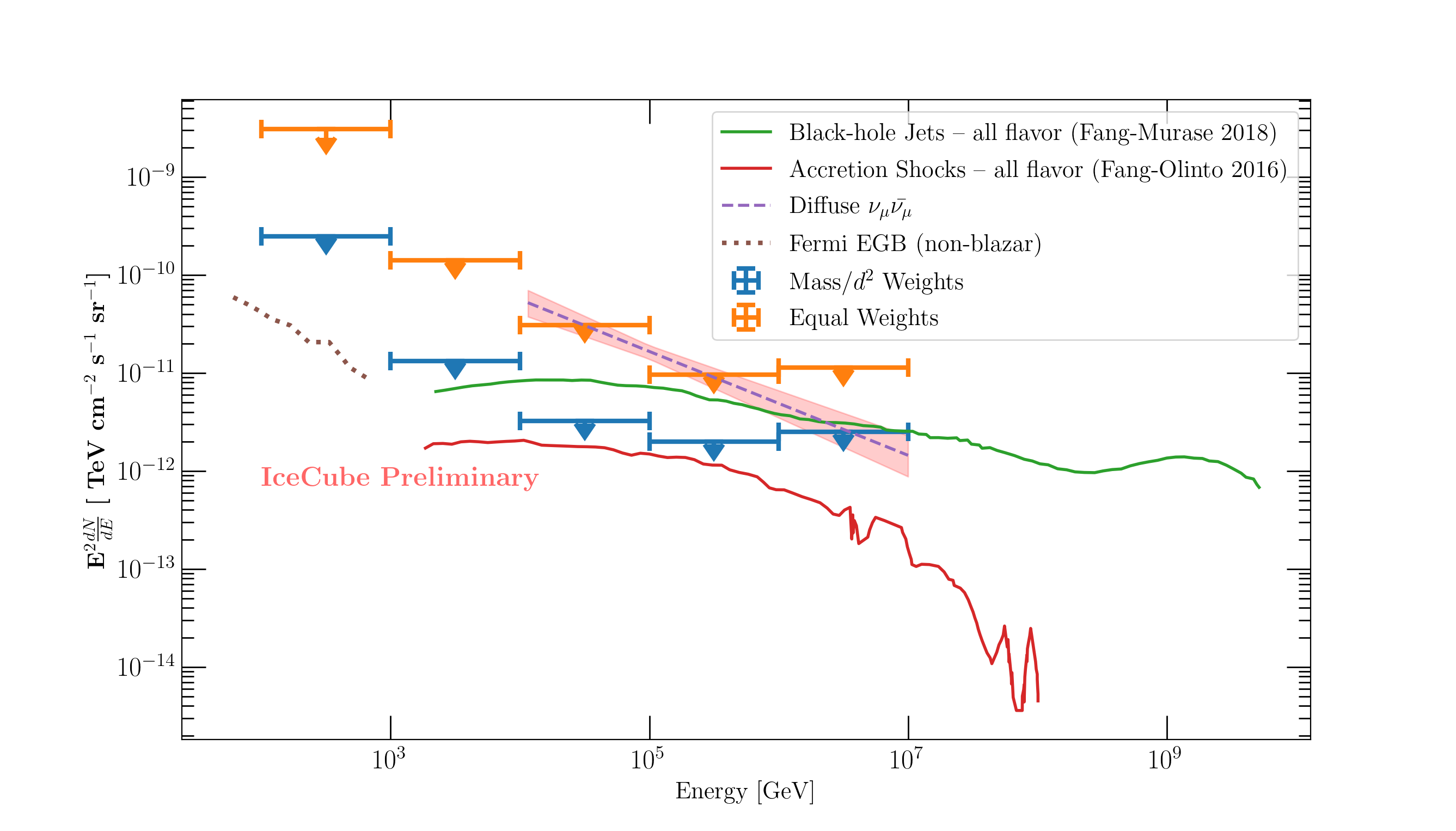}
    \caption{\label{fig:diffsens} IceCube 90\% upper-limits on the stacked flux from the Planck SZ Catalog of galaxy clusters in quasi-differential bins of width 1 in log E (GeV). The limits have been scaled for the completeness of the catalog for both weighting schemes used (equal weights and flux-proxy weights). The green solid line shows the model prediction from Ref.~\cite{fangmurase2018} in which AGN embedded in galaxy clusters accelerate cosmic rays that result in a steady flux of neutrinos through secondary interactions. The red solid line shows the model prediction from Ref.~\cite{fangolinto2016} in which accretion shocks in galaxy clusters are responsible for CR acceleration. }
\end{figure}
%%%%%%%%%%%%%%%%%%%%%%FIGURE%%%%%%%%%%%%%%%%%%%%%%%%%%%%

\begin{table*}[h!]
\centering
\begin{tabularx}{\textwidth}{| X | X | X | X |}
\hline
 Weighting Scheme & $\gamma = 2.0$ Contribution & $\gamma = 2.5$ Contribution & $\gamma = 3.0$ Contribution\\
 \hline
 Equal & 32\% & 67\% & 28.3\% \\
 Distance & 3.9\% & 5.3\% & 1.4\% \\
 Flux-Proxy & 3.7\% & 3.9\% & 1.3\%\\
 \hline
 \end{tabularx}
 \caption{The percentages of diffuse flux constrained at 100 TeV under the simple power-law assumptions with a spectral index of 2.0, 2.5 and 3.0.}
\label{tab:contrib}
\end{table*}

We also calculate the differential upper limits in decade energy bins, assuming an $E^{-2}$ spectrum across the bin.
At energies above 10~TeV, we can exclude the Fang-Murase model found in~\cite{fangmurase2018}. The differential upper limits are shown in Fig.~\ref{fig:diffsens} for two weighting schemes, along with the predicted fluxes from the aforementioned Fang-Murase model~\cite{fangmurase2018} and the Fang-Olinto model~\cite{fangolinto2016}.

\section{Conclusions}
\label{sec:conclusions}
This work presents an IceCube search for neutrinos from galaxy clusters, some of the most massive gravitationally bound large-scale structures in the universe. 
We find no evidence of significant neutrino emission from the galaxy clusters detected via the SZ effect in the Planck survey, challenging certain scenarios of neutrino production in such objects following CR confinement and interactions in the ICM. 
Depending on the assumed weighting scheme and spectral index, we constrain the contribution of galaxy clusters, with masses between $10^{14}$ \(\textup{M}_\odot\) and $10^{15}$ \(\textup{M}_\odot\) at a redshift between 0.01 and 2, to the diffuse neutrino flux between 4\% and 67\% at 100~TeV.

\bibliographystyle{ICRC}
\bibliography{references}

% \begin{thebibliography}{99}
% \bibitem{...}
% ....

% \end{thebibliography}

% Full authors list (ONLY FOR COLLABORATIONS)
\clearpage
\section*{Full Author List: IceCube Collaboration}

\scriptsize
\noindent
R. Abbasi$^{17}$,
M. Ackermann$^{59}$,
J. Adams$^{18}$,
J. A. Aguilar$^{12}$,
M. Ahlers$^{22}$,
M. Ahrens$^{50}$,
C. Alispach$^{28}$,
A. A. Alves Jr.$^{31}$,
N. M. Amin$^{42}$,
R. An$^{14}$
% ...
\\

\noindent
% ...
$^{12}$ Universit{\'e} Libre de Bruxelles, Science Faculty CP230, B-1050 Brussels, Belgium \\
% ...
$^{14}$ Department of Physics and Laboratory for Particle Physics and Cosmology, Harvard University, Cambridge, MA 02138, USA \\
% ...
$^{17}$ Department of Physics, Loyola University Chicago, Chicago, IL 60660, USA \\
$^{18}$ Dept. of Physics and Astronomy, University of Canterbury, Private Bag 4800, Christchurch, New Zealand \\
% ...
$^{22}$ Niels Bohr Institute, University of Copenhagen, DK-2100 Copenhagen, Denmark \\
% ...
$^{28}$ D{\'e}partement de physique nucl{\'e}aire et corpusculaire, Universit{\'e} de Gen{\`e}ve, CH-1211 Gen{\`e}ve, Switzerland \\
% ...
$^{31}$ Karlsruhe Institute of Technology, Institute for Astroparticle Physics, D-76021 Karlsruhe, Germany  \\
% ...
$^{42}$ Bartol Research Institute and Dept. of Physics and Astronomy, University of Delaware, Newark, DE 19716, USA \\
% ...
$^{50}$ Oskar Klein Centre and Dept. of Physics, Stockholm University, SE-10691 Stockholm, Sweden \\
% ...
$^{59}$ DESY, D-15738 Zeuthen, Germany \\
% ...

\subsection*{Acknowledgements}

\noindent
USA {\textendash} U.S. National Science Foundation-Office of Polar Programs,
U.S. National Science Foundation-Physics Division,
U.S. National Science Foundation-EPSCoR,
Wisconsin Alumni Research Foundation,
Center for High Throughput Computing (CHTC) at the University of Wisconsin{\textendash}Madison,
Open Science Grid (OSG),
Extreme Science and Engineering Discovery Environment (XSEDE),
Frontera computing project at the Texas Advanced Computing Center,
U.S. Department of Energy-National Energy Research Scientific Computing Center,
Particle astrophysics research computing center at the University of Maryland,
Institute for Cyber-Enabled Research at Michigan State University,
and Astroparticle physics computational facility at Marquette University;
Belgium {\textendash} Funds for Scientific Research (FRS-FNRS and FWO),
FWO Odysseus and Big Science programmes,
and Belgian Federal Science Policy Office (Belspo);
Germany {\textendash} Bundesministerium f{\"u}r Bildung und Forschung (BMBF),
Deutsche Forschungsgemeinschaft (DFG),
Helmholtz Alliance for Astroparticle Physics (HAP),
Initiative and Networking Fund of the Helmholtz Association,
Deutsches Elektronen Synchrotron (DESY),
and High Performance Computing cluster of the RWTH Aachen;
Sweden {\textendash} Swedish Research Council,
Swedish Polar Research Secretariat,
Swedish National Infrastructure for Computing (SNIC),
and Knut and Alice Wallenberg Foundation;
Australia {\textendash} Australian Research Council;
Canada {\textendash} Natural Sciences and Engineering Research Council of Canada,
Calcul Qu{\'e}bec, Compute Ontario, Canada Foundation for Innovation, WestGrid, and Compute Canada;
Denmark {\textendash} Villum Fonden and Carlsberg Foundation;
New Zealand {\textendash} Marsden Fund;
Japan {\textendash} Japan Society for Promotion of Science (JSPS)
and Institute for Global Prominent Research (IGPR) of Chiba University;
Korea {\textendash} National Research Foundation of Korea (NRF);
Switzerland {\textendash} Swiss National Science Foundation (SNSF);
United Kingdom {\textendash} Department of Physics, University of Oxford.

% \noindent \textbf{Note comment afterwards:} Collaborations have the possibility to provide an authors list in xml format which will be used while generating the DOI entries making the full authors list searchable in databases like Inspire HEP. For instructions please go to icrc2021.desy.de/proceedings or contact us under icrc2021proc@desy.de.\\

% \scriptsize
% \noindent
% first.author$^1$, 
% second.author$^2$, 
% third.author$^3$ % .... more names
% and 
% last.author$^{n}$ \\

% \noindent
% $^1$first.affiliation.
% $^2$second.affiliation. % .... more affiliation
% $^{m}$last.affiliation.

\end{document}